\documentclass[a4paper,fleqn,usenatbib]{mnras}
\usepackage{txfonts}
\usepackage[T1]{fontenc}
\usepackage{ae,aecompl}
\usepackage{graphicx}    
\usepackage{color,ulem}
\usepackage{cancel}
\usepackage{slashed}
\usepackage{lipsum}
\usepackage{float}
\usepackage{relsize}
\usepackage{color}
\usepackage{xcolor}
\usepackage{multirow}
\usepackage{hyperref}

\def\ll{\left}
\def\rr{\right}

\def\Msun{M_\odot}

\def\dg{^\circ}

\def\wrt{\textit{w.r.t }}
\def\thatis{\textit{ie, }}
\def\etc{\textit{etc. }}

\def\Rmerg{R_{merger}}
\def\rdet{r_{det}}

\def\Vsens{\ll<V\rr>_{det}}
\def\VT{\ll<VT\rr>_{det}}
\def\Vmax{V_{max}} 

\def\Ndet{N_{det}}
\def\Nmax{N_{max}}
\def\rth{\rho_{th}}
\def\rthNet{\rho_{th}^{net}}
\def\rthSingle{\rho_{th}^{single}}

\def\OthreeRateUntrigFullMin{4.8}
\def\OthreeRateUntrigFullMed{5.7}
\def\OthreeRateUntrigFullMax{13.7}
\def\OthreeRateUntrigOnaxisMin{0.9}
\def\OthreeRateUntrigOnaxisMed{1.1}
\def\OthreeRateUntrigOnaxisMax{2.6}
\def\OthreeRateUntrigOffaxisMin{3.9}
\def\OthreeRateUntrigOffaxisMed{4.6}
\def\OthreeRateUntrigOffaxisMax{11.1}
\def\OthreeRateTotalFullMin{5.6}
\def\OthreeRateTotalFullMed{6.5}
\def\OthreeRateTotalFullMax{15.9}
\def\OthreeRateTotalOnaxisMin{1.5}
\def\OthreeRateTotalOnaxisMed{1.7}
\def\OthreeRateTotalOnaxisMax{4.1}
\def\OthreeRateTotalOffaxisMin{4.1}
\def\OthreeRateTotalOffaxisMed{4.8}
\def\OthreeRateTotalOffaxisMax{11.8}
\def\OthreeJointRateFullMin{1.9}
\def\OthreeJointRateFullMed{2.2}
\def\OthreeJointRateFullMax{5.5}
\def\OthreeJointRateOnaxisMin{1.1}
\def\OthreeJointRateOnaxisMed{1.3}
\def\OthreeJointRateOnaxisMax{3.2}
\def\OthreeJointRateOffaxisMin{0.8}
\def\OthreeJointRateOffaxisMed{0.9}
\def\OthreeJointRateOffaxisMax{2.3}

\def\DesignedRateUntrigFullMin{20.1}
\def\DesignedRateUntrigFullMed{23.5}
\def\DesignedRateUntrigFullMax{57.3}
\def\DesignedRateUntrigOnaxisMin{3.7}
\def\DesignedRateUntrigOnaxisMed{4.3}
\def\DesignedRateUntrigOnaxisMax{10.5}
\def\DesignedRateUntrigOffaxisMin{16.4}
\def\DesignedRateUntrigOffaxisMed{19.2}
\def\DesignedRateUntrigOffaxisMax{46.7}
\def\DesignedRateTotalFullMin{22.8}
\def\DesignedRateTotalFullMed{26.7}
\def\DesignedRateTotalFullMax{64.8}
\def\DesignedRateTotalOnaxisMin{5.8}
\def\DesignedRateTotalOnaxisMed{6.8}
\def\DesignedRateTotalOnaxisMax{16.4}
\def\DesignedRateTotalOffaxisMin{17.0}
\def\DesignedRateTotalOffaxisMed{19.9}
\def\DesignedRateTotalOffaxisMax{48.4}
\def\DesignedJointRateFullMin{6.9}
\def\DesignedJointRateFullMed{8.1}
\def\DesignedJointRateFullMax{19.6}
\def\DesignedJointRateOnaxisMin{4.3}
\def\DesignedJointRateOnaxisMed{5.1}
\def\DesignedJointRateOnaxisMax{12.3}
\def\DesignedJointRateOffaxisMin{2.6}
\def\DesignedJointRateOffaxisMed{3.0}
\def\DesignedJointRateOffaxisMax{7.3}

\def\AplusRateUntrigFullMin{140.7}
\def\AplusRateUntrigFullMed{164.8}
\def\AplusRateUntrigFullMax{400.6}
\def\AplusRateUntrigOnaxisMin{26.5}
\def\AplusRateUntrigOnaxisMed{31.0}
\def\AplusRateUntrigOnaxisMax{75.4}
\def\AplusRateUntrigOffaxisMin{114.2}
\def\AplusRateUntrigOffaxisMed{133.8}
\def\AplusRateUntrigOffaxisMax{325.3}
\def\AplusRateTotalFullMin{152.2}
\def\AplusRateTotalFullMed{178.4}
\def\AplusRateTotalFullMax{433.4}
\def\AplusRateTotalOnaxisMin{37.6}
\def\AplusRateTotalOnaxisMed{44.1}
\def\AplusRateTotalOnaxisMax{107.1}
\def\AplusRateTotalOffaxisMin{114.6}
\def\AplusRateTotalOffaxisMed{134.3}
\def\AplusRateTotalOffaxisMax{326.3}
\def\AplusJointRateFullMin{29.6}
\def\AplusJointRateFullMed{34.6}
\def\AplusJointRateFullMax{84.3}
\def\AplusJointRateOnaxisMin{25.9}
\def\AplusJointRateOnaxisMed{30.3}
\def\AplusJointRateOnaxisMax{73.7}
\def\AplusJointRateOffaxisMin{3.7}
\def\AplusJointRateOffaxisMed{4.3}
\def\AplusJointRateOffaxisMax{10.6}

\def\gwbns{GW170817}

\graphicspath{{../figures/}}

\makeatletter

\newcommand{\Rmnum}[1]{\expandafter\@slowromancap\romannumeral #1@}
\makeatother

\title[{ BNS mergers and short-GRBs with Gaussian structured jets}]{Prospects of  joint detections of neutron star mergers and short-GRBs with Gaussian structured jets}

\author[M. Saleem.]{
M. Saleem\thanks{E-mail: msaleem@cmi.ac.in, saleem.muhammed.c@gmail.com}
\\
Chennai Mathematical Institute, Siruseri, 603103 Tamilnadu.\\
}

\date{Accepted XXX. Received YYY; in original form ZZZ}

\pubyear{2019}

\begin{document}
\label{firstpage}
\pagerange{\pageref{firstpage}--\pageref{lastpage}}
\maketitle

\begin{abstract}
	{ GW170817  was the first ever joint detection of gravitational waves (GW) from a binary neutron star (BNS) merger with the detections of short $\gamma$-ray burst (SGRB) counterparts. Analysis of the multi-band afterglow observations of over more than a year revealed that the outflow from the merger end-product was consistent with structured relativistic jet models with the core of the jet narrowly collimated to half opening angles $\sim5\dg$. 
	In this work, assuming all the BNS mergers produce Gaussian structured jets with properties as inferred for \gwbns, we explore the prospects of joint detections of BNS mergers and prompt $\gamma-$ray emission, expected during the current and upcoming upgrades of LIGO-Virgo-KAGRA detectors.  We discuss three specific observational aspects: 1) the distribution of detected binary inclination angles 2) the distance reach and 3) the detection rates.
	Unlike GW-only detections, the joint detections are greatly restricted at large inclination angles, due to the structure of the jets. We find that at lower inclination angles (say below 20$\dg$), the distance reach as well as the detection rates of the joint detections are limited by GW detectability while at larger inclinations (say above 20), they are limited by the $\gamma$-ray detectability.}
\end{abstract}


\section{Introduction}

Historic detection of a BNS merger GW170817  by LIGO/Virgo detectors \citep{GW170817} and the consequent follow up of the event in various electromagnetic (EM) frequency bands and by neutrino observatories dawned a new era in multi-messenger astronomy \citep{MMApaper,Goldstein:2017mmi,Savchenko:2017ffs,KN-discovery-170817,Evans:2017mmy,Kasliwal:2017ngb,Hallinan2017b,Pian:2017gtc,Ruan:2017bha,Lyman:2018qjg,Margutti:2018xqd,DAvanzo:2018zyz,Troja:2018ruz,Resmi:2018wuc,Dobie:2018zno,Nynka:2018vup,Alexander:2018dcl,Piro:2018bpl,Lamb:2018qfn}. Multimessenger observations provide tremendous opportunities to investigate astrophysics, cosmology, and fundamental physics, for example, the estimation     of Hubble constant \citep{H0:joint-LVC-DES,H0:Hotokezaka:2018dfi}, testing the speed of gravity \citep{GRB+GW-2017}, estimating the neutron star EOS \citep{Coughlin:2018miv,Coughlin:2018fis,Radice:2018ozg} are a few among them. With more GW detectors expected to be operational~\citep{AdLIGO-2015-Aasi,Virgo-2015Acernese,Kagra-design,LIGO-India} with the improved sensitivities, we expect tens to hundreds of BNS mergers and tens of joint BNS-EM detections over next few years \citep{OSD:2013wya,Coward:2012gn,clark2015prospects,regimbau2015revisiting,Howell:2018nhu}.

One of the common parameters which has an important role in the 
GW detectability of BNS mergers as well as the EM detectability of their counterparts is the inclination angle $\iota$ of the binary which is the angle between the orbital angular momentum axis and the line of sight. The inclination angle is also known as the viewing angle $\theta_v$ of the observer\footnote[1]{The reader may note that $\iota$ is defined between $0-180\dg$ whereas $\theta_v$ is considered between $0-90\dg$ assuming that the GRB jets are bipolar. Therefore, when $\iota>90\dg$, the corresponding $\theta_v$ is taken to be $\theta_v = 180-\iota$}, as commonly referred to in GRB literature and is a key parameter in understanding the physics of EM counterparts
\citep{arun2014synergy,Lamb:2017xti} especially when they are observed off-axis \citep{Granot:2002za,Donaghy:2005nq,lazzati2016off,Kathirgamaraju:2017igg,Eichler:2018nfe}. For example, the fact that GRB170817A was viewed off-axis has been central in understanding the underlying jet structure models \citep{Lamb:2017ybq,Ioka:2017nzl,Granot:2017tbr,Lazzati:2017zsj,Resmi:2018wuc,Gill:2018kcw,Ghirlanda:2018uyx,Troja-2018-SJ,Lamb:2018qfn}. The inclination angle of GW170817 was estimated independently from GW and EM observations as well as from the combined GW+EM analyses \citep{Mandel:2017fwk,Finstad:2018wid,Mooley:2018dlz}. From the observation of the superluminal motion of the radio counterpart, the inclination angle of GW170817  was constrained to $\sim [14\dg,28\dg]$ \citep{Mooley:2018dlz,Ghirlanda:2018uyx}.

The dependence of GW detectability of BNS on the inclination angle has been well formulated in literature. \citet{Schutz2011} showed that the detected inclinations follow the probability distribution $P_{det}(\iota) = 0.076076\,(1 + 6 \cos^2{\iota} + \cos^4{\iota})^{3/2} \sin{\iota}$. \cite{Seto:2014iya} also carried out a similar and detailed study on inclination angle distributions with broadly similar conclusions. 
As far as the EM counterparts are concerned, due to the collimated jets and relativistic beaming \citep{Rhoads:1999wm, Sari:1999mr, Harrison:1999hv}, their detectabilities at various frequency bands (short-GRB (SGRB) prompt $\gamma-$ray emission, afterglows \etc) have even stronger dependence on the inclination angle \citep{Granot:2002za,Yamazaki:2003af,Donaghy:2005nq,Lamb:2017ych,saleem2017agparameterspace,Ioka:2019jlj}
Unlike GW, the EM detectability dependence on inclination angle is only known subject to the assumption of an underlying model of jet structure \citep{Donaghy:2005nq,salafia2015structure} while our knowledge about the structure is still developing. The accurate knowledge of the statistical distribution of inclination angle is important in forecasting the future multimessenger detection rates and detection scenarios.  
{(See also \cite{Kochanek:1993mw,nissanke2010exploring} and \cite{saleem-etal-2017-agRates} for further discussions about the prospects of GRB beaming and joint detections.)}

The probability distributions given by \cite{Schutz2011} and \cite{Seto:2014iya} are excellent representations for the detected BNS inclinations from GW observations alone. However, in a multimessenger observing scenario, due to the BNS-SGRB  association, it is possible to have SGRB triggered BNS detections which can be made at a lower detection threshold than needed for an independent GW detection. This brings in several additional BNS detections. Due to the strong inclination angle dependence of SGRB detectability, these additional detections also depend on the inclinations. Consequently, the distribution of overall (independent plus SGRB triggered ) detected  BNS inclinations will be different from the ones given by \citep{Schutz2011,Seto:2014iya}. Apart from these two distributions, one would also be interested in the distribution for joint (simultaneous) detections of BNS and SGRBs. 

The subject of this work is to demonstrate these new inclination distributions and to discuss { how the detection rates and distance reach depend on the inclinations}.  We use a Monte Carlo simulated population of BNS mergers and associated SGRBs to compute the inclination angle distributions for independent BNS detections, SGRB-triggered BNS detections, and joint BNS-SGRB detections. To obtain further insights, we compute the detection rates for low and high inclination angle cases and compute the distance reaches as a function of inclination angle. We investigate how these properties vary if the sensitivity of the GW detector network increases, by considering the current and upcoming networks with three, four and five detectors and their projected sensitivities.

In section \ref{sec2}, we present our simulated population and obtain the inclination angle distributions and also compute the distance reaches. In section \ref{sec3}, we discuss the detection rates and in section \ref{sec4} we conclude the work.

\begin{figure*}
	\includegraphics[scale=0.9]{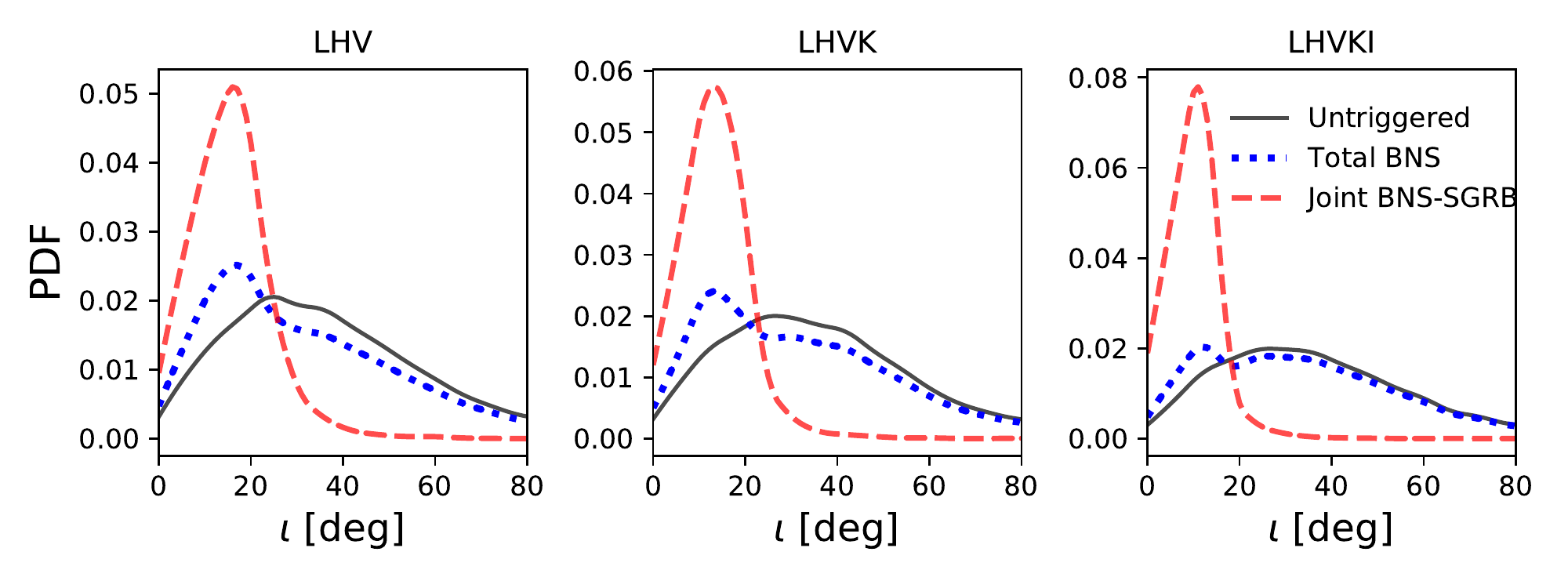}
	\caption{Probability distribution of inclination angles for \textit{untriggered} BNS detections (solid curves), total BNS detections (\textit{triggered} plus \textit{untriggered}) (dotted curves) and the \textit{joint} BNS-SGRB detections (dashed curves) for various network configurations. The curves are independently normalized such that the area under each curve is unity.}
	\label{fig:rate-vs-iota}
\end{figure*}

\section{Simulated BNS-SGRB populations and Inclination angle distributions}
\label{sec2}

In this section, we numerically compute the distributions of detected inclination angles of BNS mergers for multimessenger observing scenarios. By multimessenger observations, we precisely mean the GW detections of BNS mergers together with the EM detections of SGRB prompt $\gamma-$ray emissions. EM counterparts in relatively lower frequencies such as afterglows and  Kilonovae emissions are not considered. This is because the prompt emission provides the best opportunity for an EM-triggered BNS detection due to their strong temporal coincidence with BNS merger (delay at the order of seconds) while afterglow or kilonovae emission appears with a delay of the order of hours to days or even longer \citep{metzger2012most}. 

\subsection{Simulation of BNS merger population}    
We simulate a population of $10^6$ BNS mergers and associated SGRB prompt emission counterparts. The component NS masses are distributed uniformly between $1-2\Msun$ and component spins are taken to be zeros. {The non-spinning approximation is motivated by} the fact  that the fastest  spinning NS in a binary system {among the 18 galactic BNS systems} known till date  has a dimensionless spin magnitude {$\chi < 0.05$} \citep{Zhu:2017znf} and the effects of such spins  on the BNS detectability are only moderate \citep{Brown:2012qf}. Sources are distributed in comoving volume with constant number density and the source orientation is taken as uniform over the polarisation sphere, \textit{ie}, {the in-plane binary orientation $\psi$ is distributed uniformly in [0,$2\pi$] and the cosine of inclination ($\cos\iota$) is distributed uniformly in [0,1].} 

For all the simulated sources, we computed the single-detector SNRs at each detector as well as the coherent network SNR \citep{Schutz2011} for various network configurations discussed below. 

\begin{enumerate}
	
	\item  
	\textit{LHV} - the three-detector LIGO-Virgo network  with the projected O3 sensitivities for two LIGO detectors and the best reported O2 sensitivity \citep{gwtc-1} for Virgo which we consider as a conservative choice for O3 sensitivity. This configuration can be considered to be representative of the ongoing observation run O3 (during 2019-2020). 
	
	\item 
	\textit{LHVK} - the four-detector network with the addition of KAGRA\citep{Kagra-design} with all of them at their designed sensitivities \citep{AdLIGO-Harry-2010,AdLIGO-2015-Aasi,Virgo-2015Acernese}. This can be considered as the representative configuration for the near-future years (2021-2023)
	
	\item  
	\textit{LHVKI} - the five-detector network with the further addition of LIGO India \citep{LIGO-India}.  All the three LIGO detectors are considered at the Aplus sensitivities \citep{aplus:Barsotti} while Virgo and Karga are considered at their designed sensitivities. LIGO India is expected to be operational around 2024-25 and hence this configuration may represent the epoch of 2024-2026. {Refer to the URL given in the footnote}\footnote[2]{Sensitivity curves are taken from the URL: \url{https://dcc.ligo.org/LIGO-T1500293/public}} { for all the sensitivity curves.}
\end{enumerate}

Note that for each network configuration, the sources are distributed up to a distance which is the farthest detectable distance (Horizon distance) for that configuration. 

\subsection{SGRB counterparts with Gaussian structured jets}
	For each BNS in the population, its SGRB counterpart is considered to have structured jets \citep{Rossi:2001pk,Zhang_2002,Nakar_2004,Lamb:2017ych} with Gaussian-like profiles. In a Gaussian structured jet \citep{Zhang_2002,salafia2015structure,Resmi:2018wuc}, the energy and bulk Lorentz factor in the source frame follow Gaussian-like  angular variations, $\propto \exp{[-(\theta/\theta_c)^2]}$, where $\theta$ is the angle away from the jet axis and $\theta_c$ is {1$\sigma$} width of the Gaussian profile which is also referred to as the semi-opening angle of the core of the Gaussian jet. 

	It was found that the observed features of EM counterparts of GW170817 (prompt $\gamma-$rays and afterglows) were consistent with  Gaussian structured jets
	{  \citep{Resmi:2018wuc,Lamb:2017ybq,Howell:2018nhu,Lamb:2018qfn} with $\theta_c \leq 6\dg$ and}
	viewed far off-axis, \thatis $> 20\dg$ (Recall that the best estimate of $\theta_v$ obtained from VLBI observations is around $20\dg$ \citep{Mooley:2018dlz}). In our simulated population, we assume that all the BNS mergers have jet counterparts with properties consistent with what is inferred for GRB170817A, with $\theta_c=5\dg$. This is consistent with a recent study by \cite{Beniamini:2018uwo} which argues that most neutron star mergers result in tightly collimated successful jets.  
	For all the sources in the population, we also assume the initial bulk Lorentz factor at the axis of the jet $\Gamma(\theta_v=0) = 100$ and  {$E_{\gamma}=10^{49}$} erg, where $E_{\gamma}$ is the rest frame total $\gamma$-ray energy. 

	To estimate the detectability of prompt $\gamma-$ray emission, it also requires to compute the observable quantities such as $\gamma-$ray flux (in units of erg cm$^{-2}$ s$^{-1}$) or fluence (in units of erg cm$^{-2}$) in the sensitive bandwidth of the instrument under consideration. In a most recent work, \cite{sreelakshmiEtAl} has numerically computed the prompt $\gamma-$ray fluence for Fermi GBM assuming Gaussian structured jet with parameters as described above, where the fluence is obtained for a range of viewing angles from 0 to 90$\dg$, at a luminosity distance $D_L = 41$ Mpc which is the inferred luminosity distance of GW170817. In order to estimate the detectability of SGRB counterparts of the BNS mergers in our population, we use this fluence data by scaling appropriately for desired luminosity distances. If the fluence obtained in \cite{sreelakshmiEtAl} is denoted as $f_{\gamma,41Mpc}(\theta_v)$, then the fluence from a source at an arbitrary distance $D_L$ at viewing angle $\theta_v$ can be schematically written by the scaling relation,   
	\begin{equation}
		f_{\gamma}(\theta_v,D_L) = \ll(\frac{41Mpc}{D_L}\rr)^2\,f_{\gamma,41Mpc}(\theta_v).
	\label{eq:fluence}
	\end{equation}
	
\subsection{Detectabilities}    
	Given the population of BNS mergers along with their SGRB counterparts described above, we now asses their detectabilities by imposing appropriate detection thresholds on their SNR and fluence respectively. We consider the following two scenarios: 

\begin{enumerate}
	
	\item 
		BNS mergers are detected from independent GW observations. For this case, we use a network SNR threshold $\rthNet = 10$ along with a coincident single detector threshold $\rthSingle = 4$ at minimum two detectors as the detection criterion. Henceforth, these are referred to as \textit{untriggered} detections as these are not triggered by observations in any other window.
	
	\item 
		BNS mergers are detected from the SGRB-triggered search. Here the SGRB must have been detected either independently or by following up the low significant BNS candidates (which do not qualify itself as significant detections). For this, we use network SNR threshold $\rthNet = 8$ along with a coincident single detector threshold $\rthSingle = 3$ at minimum two detectors as the detection criterion. 
\end{enumerate}

	The lowered SNR thresholds ($\rthNet=8$ and $\rthSingle = 3$) for the SGRB-triggered searches follow from the fact that the presence of a detected SGRB counterpart allows targetted BNS search in a small fraction of sky over a shorter segment of time as opposed to the all-sky searches carried out over longer segments of time. This significantly reduces the number of templates in the search template bank which in turn reduces the number of background (terrestrial) triggers and hence the false alarm rates (FAR). A low FAR allows claiming significant detections at lower SNR thresholds without any cost of false detections. {This has been comprehensively discussed in a number of earlier works in literature \citep{Ilya-triggered-2011,baret2012multimessenger,Ilya:triggeredSearch,Chen:2012qh,Dietz:2012qp,clark2015prospects,2015PhRvL.115w1101B,patricelli2016prospects,Howell:2018nhu}. Our choice of thresholds with a 20\% reduction for triggered cases, as opposed to untriggered cases, is broadly consistent with the previous studies (e.g. \cite{Chen:2012qh,Dietz:2012qp,clark2015prospects}).
	In recent years, there were triggered searches performed on LIGO-Virgo data following the detections of short and long GRBs \citep{triggered-gw-search-lgrb:2013cya,triggered-gw-IPN:2014iia,triggered-gw-method-Aasi:2014ent,Abbott:2016cjt,LVC-triggered-2019}. }

	Whether a given BNS in the population belongs to the  SGRB-triggered category or the untriggered category is determined based on whether the prompt $\gamma-$ray fluence obtained by equation \ref{eq:fluence} is above the Fermi GBM detection limit. We use Fermi GBM 64-ms fluence threshold f$_{GBM}= 2\times10^{-7}$ erg cm$^{-2}$ \citep{Goldstein:2017mmi}.

\subsection{Distribution of detected inclinations}
	Applying the detectability criterion above, we obtain the following three distributions of detected BNS mergers:
\begin{enumerate}
	\item 
		\textit{Untriggered} BNS merger detections obtained by applying the SNR threshold $\rth = 10$ to all the events. 
	\item 
		\textit{Total} BNS merger detections which include both triggered and untriggered detections where the thresholds  $\rth=8$ and $\rth=10$ are appropriately applied based on the SGRB detectability.
	\item 
		\textit{Joint} BNS-SGRB detections which include only those events for which both BNS and SGRB are simultaneously detectable. This is obtained from the \textit{Total} BNS detections (above item) by removing those events for which the SGRB is not detectable.  
\end{enumerate}  
	Figure-\ref{fig:rate-vs-iota} shows the distribution of detected inclination angles corresponding to \textit{untriggered} detections (black solid curves), \textit{total} detections (blue dotted curves) and the \textit{joint} BNS-SGRB detections (red dashed curves). 
	All the distributions are independently normalized such that the area under each curve is unity.  The three panels correspond to the three different network configurations. The solid curves are the distributions corresponding to the detections from GW observations alone and hence they are very close to the analytical prediction by \cite{Schutz2011} which was mentioned earlier.

	It is observed that the solid and dotted curves (untriggered and total detections respectively) differ from each other at lower inclinations and as the inclination increases, the difference between them decreases and tend to join together. This implies that the contribution from SGRB-triggered detections mostly happens at lower inclination angles while there are hardly any additions at higher inclinations. Also,     the dashed red curves (corresponding to the joint BNS-SGRB detections), unlike other two curves, declines rapidly around $\iota$ between 20 and 30, indicating that the joint detections are extremely less likely at larger inclinations. All these features are the consequences of the fact that the prompt emission detection (and hence the `SGRB-triggering') is most likely to happen at lower inclination angles which is also a consequence of the narrowly collimated jets ($\theta_c=5\dg$ in our case) and the Lorentz boosted $\gamma-$ray emissions. 

\begin{figure}
	\centering
	\includegraphics[scale=0.7]{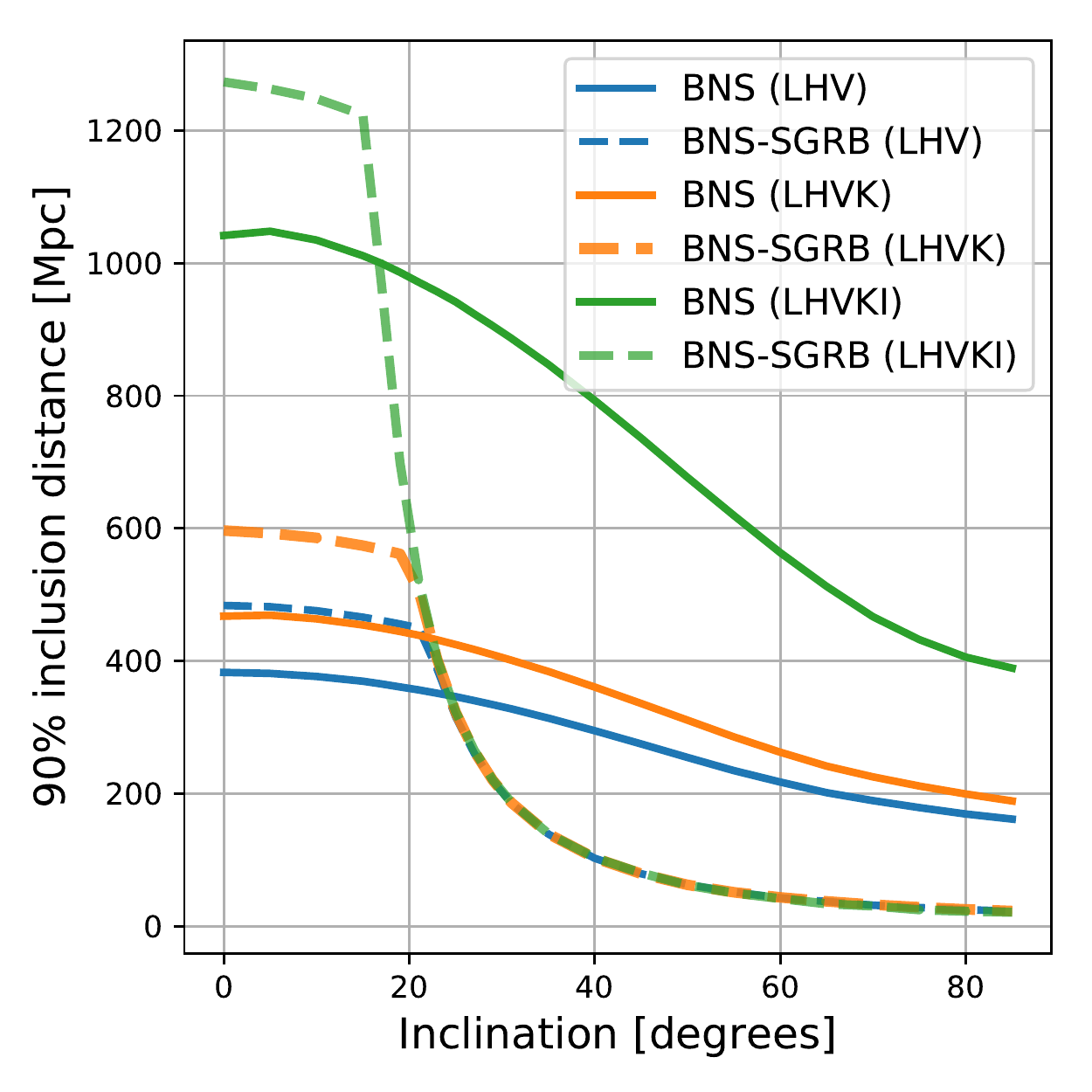}
	\caption{The 90\% inclusion distance of BNS and BNS-SGRB joint detections as a function of inclination angle. 90\% inclusion distance is the luminosity distance which includes 90\% of the detections for a source population which is uniformly distributed in comoving volume. The solid curves correspond to the BNS merger detections by GW detector networks while the dashed curves correspond to the joint (simultaneous) detections of BNS and SGRB. The three colors show the three different network configurations LHV, LHVK, and LHVKI with respective sensitivities. SGRB counterparts are assumed to have Gaussian structured jets producing prompt $\gamma-$ray emissions whose detectability were assessed for Fermi GBM detection limits.}
	\label{fig2}
\end{figure}

	Further, we notice that as the GW detector network sensitivity (or the distance reach) increases, at larger inclinations, the contributions from \textit{triggered} detections becomes less relevant. For example, among the three panels in figure-\ref{fig:rate-vs-iota}, for the \textit{LHVKI} configuration (rightmost) which is the most sensitive one, the distributions corresponding to the untriggered and total BNS detections (solid and dotted curve respectively) deviate from each other minimally compared to the other two cases (left and middle panels) which are  relatively lower sensitive. We understand this as follows. Since we have chosen SNRs of 8 and 10 as the thresholds for \textit{triggered}  and \textit{untriggered} detections, all the triggered detections must come from a subset of sources whose SNRs lie between 8 and 10. Due to the fluence dependency on $\iota$, at larger $\iota$ (say above 20$\dg$) SGRBs  can be detected only if they are at sufficiently close distances which means that SGRB-triggered BNS detections are also possible within this smaller distance limit (say at the level of 100Mpc or slightly above or below - see figure \ref{fig2} for distance reach estimates for various cases). Within this distance, with the \textit{LHVKI} configuration, almost all the sources must have got high SNR (SNR > 10) and hence there would hardly be any candidate as an SGRB-triggered detection. On the other hand, at relatively low sensitive configurations (\textit{LHV} and \textit{LHVK}), there could still be a reasonable fraction of sources with SNR between 8 and 10 and hence can be candidates for triggered detections. This shows that triggered detections become less relevant at higher inclinations as the GW sensitivity increases.     However, for sources with low inclinations (say less than 20$\dg$), SGRBs can be detected till much larger distances (say $>1Gpc$) (See figure \ref{fig2}) and hence triggered detections can still happen at large distances. 

\subsection{Distance reach vs inclination}
	An interesting quantity which is responsible for the features observed in figure \ref{fig:rate-vs-iota} is the distance reach which is also a direct consequence of the GW or EM detector sensitivities. Here we obtain the 90\% inclusion distance for the multimessenger BNS-SGRB detections as a function of the inclination angle. At any inclination, the corresponding  90\% inclusion distance is defined as the luminosity distance such that the 90\% of the total expected detections happen below this distance. { Note that the 90\% inclusion distance is different from the horizon distance which is the distance to the farthest detectable source by a given network.}

	The results are shown in figure \ref{fig2} where the solid blue curves correspond to the BNS merger detections by GW detector networks alone (\textit{untriggered} BNS detections) while the dashed curves correspond to the joint (simultaneous) detections of BNS and SGRB. We observe the following points in figure \ref{fig2}.
\begin{enumerate}
	\item 
		The \textit{untriggered} BNS distance reaches (solid curves) are continuous curves with a decreasing trend as a function of the inclination. This trend is clearly due to the fact that the GW signal amplitudes of the plus and cross polarisations scale down with inclination as $(1+\cos^2\iota)/2$ and $\cos\iota$ respectively \citep{Schutz2011}. 
	
	\item 
		The joint detection distance reaches (dashed curves) have discontinuities roughly around $\iota = 20\dg$ for all three network configurations. The first part of the curve (left to the discontinuity)  follows the  BNS reach as the SGRB detections at this range of inclinations are possible even from $> 1Gpc$. The second part of the curve (the sharp declining part to the right of the discontinuity) follows the SGRB distance reach which in turn is due to the combined effect of the assumed jet properties (structure as well as the specific properties such as $E_{\gamma},\theta_c$ \etc) and the Fermi GBM threshold. These dashed curves can be seen to be joining together which is due to the common Fermi GBM sensitivity assumed in this study alongside all the three GW network configurations. 
	
	\item 
		On the left part of the discontinuity, the reaches of joint detections (dashed curves) are deeper than the corresponding reaches of \textit{untriggered} BNS detections (solid curves). This is due to the respective choices of SNR thresholds while this is not reflected on the right to the discontinuity as there the horizon is set solely by the $\gamma$-ray detectability.  
		
	\item 
	    {Apart from the \textit{untriggered} and \textit{joint-detection} distance reaches, we have the distance reaches for the overall (total) BNS detections (equivalent to the dotted curves in figure \ref{fig:rate-vs-iota}). 
	    These are not explicitly shown in figure \ref{fig2} as they  will simply follow the dominant trend on either side of the intersect for each detector combination.}     
\end{enumerate}

\section{Detection rates and other implications}\label{sec3}

	In this section, we compute the detection rates of BNS events as well as BNS-SGRB coincident events for various network configurations. Motivated by the discontinuity features observed in Figures \ref{fig:rate-vs-iota} and \ref{fig2}, we compute detection rates for the arbitrary ranges $\iota\leq20\dg$ and $\iota>20\dg$. To compute the rates using our simulated population, we follow the method described below. 

	The BNS detection rates are obtained as the product of the intrinsic BNS merger rate density $\Rmerg$ and the comoving detection volume $\Vsens$,   
	\begin{equation}
		\rdet = \Rmerg \times \Vsens,
		\label{eq-rates}
	\end{equation}      
	where $\Rmerg$ represents the non-evolving BNS merger rate density (in units of $Gpc^{-3} yr^{-1}$). In this work, we use the merger rate density $\Rmerg = 662^{+1609}_{-565}$ estimated from O1/O2 observations by the \texttt{GstLAL} search pipeline \citep{gwtc-1}, $\Vsens$  is the detection volume of the detector network \thatis  the volume (in $Gpc^{3}$) which our detector network is sensitive to for BNS mergers. $\Vsens$ depends on the network configuration as well as their sensitivity and the properties of the population which is assumed (for example, the mass distribution which in this study is assumed to be uniform in $1-2\Msun$ for component masses while there are other models used in literature \citep{gwtc-1}). Given our simulated population,  $\Vsens$ can be estimated as 
	\begin{equation}
		\Vsens = \ll(\frac{\Ndet}{\Nmax}\rr) \times \Vmax.
	\end{equation}
	where $\Ndet$ is the number of detected sources out of the total $\Nmax$ sources which are distributed in a volume $\Vmax$. Note that the detection volume $\Vsens$ is an approximation to the $\VT$ used in the LIGO-Virgo rate calculation \citep{gwtc-1,population-bbh}. Since we assume the sensitivity of our network configurations to be static over a year, $\Vsens$ can well be treated as a time-averaged version of $\VT$. In obtaining the detection rates, we have assumed a 50\% duty cycle for each of the GW detectors and a 60\% sky coverage (time-averaged) for Fermi GBM telescope following \cite{Burns_2016}. 

	In Table-\ref{tab-det-rates-all}, we have shown the detection rates (per year) for untriggered BNS, total BNS and the joint BNS-SGRB detections in all three network configurations. Different columns correspond to the overall rates, rates for  $\iota \leq 20\dg$ and  the rates for $\iota >20\dg$ respectively.

\begin{table}\centering 
	\caption {Detection rates (per year) in three different network configurations. Three columns correspond to all the detections, detections with $\iota \leq 20\dg$ and $\iota > 20\dg$. The rows corresponding to \textit{Untriggered BNS} are the independent BNS detections, \textit{Total} corresponds to \textit{Untriggered} plus the \textit{SGRB-triggered} detections while the \textit{joint BNS-SGRB} corresponds to those cases in which both are simultaneously detected irrespective of whether triggered or untriggered.}            
	\begin{tabular}{|l|c|c|c|}
		\hline        
		
		Case & Any $\iota$   & $\iota \leq 20\dg$  & $\iota > 20\dg$ \\
		\hline\hline            
		\hspace{-5mm} \textit{LHV} &&&\\
		
		Untriggered BNS 
		& ${\OthreeRateUntrigFullMed}_{-\OthreeRateUntrigFullMin}^{+\OthreeRateUntrigFullMax}$ 
		& ${\OthreeRateUntrigOnaxisMed}_{-\OthreeRateUntrigOnaxisMin}^{+\OthreeRateUntrigOnaxisMax}$
		& ${\OthreeRateUntrigOffaxisMed}_{-\OthreeRateUntrigOffaxisMin}^{+\OthreeRateUntrigOffaxisMax}$  \vspace{1mm}\\

		Total BNS 
		& ${\OthreeRateTotalFullMed}_{-\OthreeRateTotalFullMin}^{+\OthreeRateTotalFullMax}$ 
		& ${\OthreeRateTotalOnaxisMed}_{-\OthreeRateTotalOnaxisMin}^{+\OthreeRateTotalOnaxisMax}$
		& ${\OthreeRateTotalOffaxisMed}_{-\OthreeRateTotalOffaxisMin}^{+\OthreeRateTotalOffaxisMax}$  \vspace{1mm}\\  
		
		Joint BNS-SGRB
		& ${\OthreeJointRateFullMed}_{-\OthreeJointRateFullMin}^{+\OthreeJointRateFullMax}$ 
		& ${\OthreeJointRateOnaxisMed}_{-\OthreeJointRateOnaxisMin}^{+\OthreeJointRateOnaxisMax}$ 
		& ${\OthreeJointRateOffaxisMed}_{-\OthreeJointRateOffaxisMin}^{+\OthreeJointRateOffaxisMax}$  \\  
		\hline
		\hspace{-5mm} \textit{LHVK} &&&\\
		Untriggered BNS 
		& ${\DesignedRateUntrigFullMed}_{-\DesignedRateUntrigFullMin}^{+\DesignedRateUntrigFullMax}$ 
		& ${\DesignedRateUntrigOnaxisMed}_{-\DesignedRateUntrigOnaxisMin}^{+\DesignedRateUntrigOnaxisMax}$
		& ${\DesignedRateUntrigOffaxisMed}_{-\DesignedRateUntrigOffaxisMin}^{+\DesignedRateUntrigOffaxisMax}$  \vspace{1mm}\\

		Total BNS 
		& ${\DesignedRateTotalFullMed}_{-\DesignedRateTotalFullMin}^{+\DesignedRateTotalFullMax}$ 
		& ${\DesignedRateTotalOnaxisMed}_{-\DesignedRateTotalOnaxisMin}^{+\DesignedRateTotalOnaxisMax}$
		& ${\DesignedRateTotalOffaxisMed}_{-\DesignedRateTotalOffaxisMin}^{+\DesignedRateTotalOffaxisMax}$  \vspace{1mm}\\ 
		
		Joint BNS-SGRB 
		& ${\DesignedJointRateFullMed}_{-\DesignedJointRateFullMin}^{+\DesignedJointRateFullMax}$ 
		& ${\DesignedJointRateOnaxisMed}_{-\DesignedJointRateOnaxisMin}^{+\DesignedJointRateOnaxisMax}$ 
		& ${\DesignedJointRateOffaxisMed}_{-\DesignedJointRateOffaxisMin}^{+\DesignedJointRateOffaxisMax}$ \\
		\hline
		
		\hspace{-5mm} \textit{LHVKI} &&& \\
		Untriggered BNS 
		& ${\AplusRateUntrigFullMed}_{-\AplusRateUntrigFullMin}^{+\AplusRateUntrigFullMax}$ 
		& ${\AplusRateUntrigOnaxisMed}_{-\AplusRateUntrigOnaxisMin}^{+\AplusRateUntrigOnaxisMax}$
		& ${\AplusRateUntrigOffaxisMed}_{-\AplusRateUntrigOffaxisMin}^{+\AplusRateUntrigOffaxisMax}$  \vspace{1mm}\\

		Total BNS 
		& ${\AplusRateTotalFullMed}_{-\AplusRateTotalFullMin}^{+\AplusRateTotalFullMax}$ 
		& ${\AplusRateTotalOnaxisMed}_{-\AplusRateTotalOnaxisMin}^{+\AplusRateTotalOnaxisMax}$
		& ${\AplusRateTotalOffaxisMed}_{-\AplusRateTotalOffaxisMin}^{+\AplusRateTotalOffaxisMax}$  \vspace{1mm}\\ 
		
		Joint BNS-SGRB 
		& ${\AplusJointRateFullMed}_{-\AplusJointRateFullMin}^{+\AplusJointRateFullMax}$ 
		& ${\AplusJointRateOnaxisMed}_{-\AplusJointRateOnaxisMin}^{+\AplusJointRateOnaxisMax}$ 
		& ${\AplusJointRateOffaxisMed}_{-\AplusJointRateOffaxisMin}^{+\AplusJointRateOffaxisMax}$  \\

		\hline\hline
	\end{tabular}
	\label{tab-det-rates-all}
\end{table}

We make the following observations from Table \ref{tab-det-rates-all}:
\begin{itemize}
	\item 
		The fraction of total BNS detections which are joint BNS-SGRB detections is 34\% in \textit{LHV} configuration, 30\% in \textit{LHVKI}, and 19\% in \textit{LHVKI}. The decrease in joint detection fraction at higher GW sensitivity is due to the several BNS detections at large inclinations for which the SGRB counterpart is undetected.     
	
	\item      
		Due to the SGRB-triggered BNS detections, the overall BNS rate increases by 29\% (\textit{LHV}), 26\% (\textit{LHVKI}) and 8\% (\textit{LHVKI}). This shows that the contribution from triggered detections becomes less relevant as the sensitivity of the GW network increases. Also, their relevance gets restricted to the low inclination angle cases, with the improvement being above 40\% for $\iota\leq20\dg$ cases and only below 5\% for $\iota > 20\dg$ cases. This is a direct consequence of the previous point that most BNS with $\iota > 20\dg$ have their SGRB counterparts undetected.    
	
	\item     
		The number of joint detections with  $\iota \leq 20\dg$ is several times larger than the number of joint detections with $\iota > 20\dg$. For the \textit{LHV}, \textit{LHVK} and the \textit{LHVKI} configurations considered in this paper, the ratio of number of events with $\iota \leq 20\dg$ to those with $\iota > 20\dg$ are 10:7, 10:6 and 7:1 respectively.
	
	\item 
		For $\iota \leq 20\dg$, the joint detection rates are very close to the total BNS detection rates, for all the three networks. The small differences found in the table are due to the limited sky coverage of Fermi GBM, rather than sensitivity. For a 100\% sky-coverage instrument, the joint detection rates will be very close to the total BNS detection rates when $\iota < 20\dg$. In other words, in the era of the second-generation GW detectors, most of the BNS detections with $\iota < 20\dg$ are SGRB-detectable too, unless restricted by instrument's sky coverage. 
	
	\item 
		The joint BNS-SGRB detection rate for the LHV configuration is $\OthreeJointRateFullMed^{+\OthreeJointRateFullMax}_{-\OthreeJointRateFullMin}$ per year. It is worth noting that this is broadly consistent with the local short GRB rates obtained in \cite{Mandhai:2018cdl} where the upper limit to the all-sky detection rate of local short GRBs (estimated from Swift observations) is < 4 per year from within $D_L$ < 200 Mpc.	             
	
\end{itemize}

\section{Conclusion and Outlook}\label{sec4}

	In this work, we have discussed { the prospects of} multimessenger detections of BNS 
	{ mergers, assuming all BNS mergers produce relativistic structured jets with Gaussian profiles as inferred for GRB170817A.} 
	We demonstrated the { expected distributions of inclination angles for BNS detections in the untriggered, triggered and joint detection scenarios}
	and forecasted the  distance reach and detection rates of BNS detections { for the sensitivities at the current and upcoming upgrades of LIGO-Virgo-KAGRA detectors}. 
	{We find that the narrowly collimated and highly relativistic nature of the SGRB jets leads to strong correlation between the detectability of the SGRB counterpart and the binary inclination angle and as a result, the chances of joint detection diminish greatly at higher inclination angles. We have demonstrated these effects through the distance reach and detection rates showing them as a function of inclination angle.}

	{ The present analysis has considered Fermi GBM as the only $\gamma$-ray detection facility alongside the multi-detector network of detectors for GW detections. As mentioned before, Fermi GBM has $\sim 60\%$ duty cycle (the time-averaged sky coverage) and limited sky localisation ability which is of the order of several square degrees. Our results, in strict sense, are sensitive to these specific features and subject to vary if we consider other $\gamma$-ray instruments to be simultaneously operational, though the broader conclusions of the paper will remain the same. For example, INTEGRAL/SPI-ACS has an effective duty cycle of 85\% \citep{Savchenko:2017xjd} which if operated simultaneously with Fermi GBM, will result in a combined duty cycle of $\sim$94\% which in turn will increase the chances of triggered detections and joint detection. With the inclusion of Swift BAT which has only 11\% duty cycle, the combined duty cycle further increases to at most $\sim95\%$ while Fermi and Swift alone provide a combined duty cycle\footnote[3]{If $D_1$, $D_2,..D_N$ are the duty cycles normalized to unity, of N instruments in operation, then the combined duty cycle, which can be interpreted as the probability that at least one of them is observing a given event, is obtained as ${ D_{combined} = 1 - \prod_{i=1}^N (1-D_i)}$} of $\sim$64\%} 
	
	{ On the other hand, Swift has a much better ability for sky localization (order of arcmin) compared to Fermi which will enable follow-up observations in other EM bands such as kilonovae and/or afterglows which will perhaps lead to avail the redshift information either from an optical counterpart or via the identification of host galaxies. The redshift information will significantly improve the source characterization of BNS merger and will also have implications for fundamental physics such as probing GW polarization, Hubble constant measurements \etc
	As far as the results of this paper are concerned, the inclusion of additional instruments will alter the over-all detection rates and distance reaches slightly while their dependence on the binary inclination as well as the distribution of inclination angles themselves will remain pretty much the same. Again, this follows from the fact that inclination distributions  are more of a reflection of the GW emission profiles and structured jet profiles considered, rather than being affected by the sensitivity. }

	{Further, } since the inclination angle distributions of multimessenger BNS detections (dashed and dotted lines in figure \ref{fig:rate-vs-iota}) carry the imprints of the jet structure models, such distributions obtained from accumulated detections over several years can be used to probe different structured-jet models. \cite{miller2015constraining} explored a similar idea using viewing angle distributions of independent SGRB detections and found that around 300 GRBs are sufficient for distinguishing their structure if their viewing angles are reconstructed well. Viewing angle distributions from multimessenger observations (as shown in this work) carries the added advantage of being more accurately estimated due to the GW detection with a multi-detector network \citep{arun2014synergy,tagoshi2014parameter,Chen:2018omi} as obtained for GW170817 \citep{Mandel:2017fwk,Finstad:2018wid}.

	As mentioned before, we have not considered the X-ray, optical and radio counterparts for triggering BNS searches and detections, primarily due to their large time window for the coincidence \wrt the merger which makes it difficult to establish a GW-EM temporal coincidence. Moreover, independent detections of such delayed counterparts are challenging too which may be attributed to the limited field of view of many present EM instruments as well as the contamination from other EM transients such as supernovae. { Therefore, with the present-day facilities, it will be extremely rare cases to identify transients from optical or radio observations, unless there are no GW or GRB detections. } 

	{ One caveat we would like to mention about this study } is that we have assumed all the SGRBs to have GRB170817A-like properties which we invoked for simplicity in this study. In reality, the properties can vary from source to source. However, our current knowledge about the distributions of such properties such as $E_{\gamma}, \theta_c$ are very limited  

\section{Acknowledgements}
	The author is partially supported by a grant from the Infosys Foundation. The author is extremely thankful to  K.G. Arun and L. Resmi for useful discussions and inputs over the course of this entire work. {The author is grateful to the anonymous referee for very useful comments and suggestions which helped to improve the quality of the manuscript significantly.} The author also thanks K. Haris, B. Sathyaprakash and N.V. Krishnendu for useful comments and inputs and thanks to Jilva for proofreading. This document has LIGO preprint number \texttt{P1900134}.

\bibliographystyle{mnras}
\bibliography{draft}

\end{document}